\documentclass[twocolumn,showpacs,preprintnumbers,amsmath,amssymb,aps]{revtex4-1}
\usepackage{graphicx}
\usepackage{dcolumn}
\usepackage{bm}
\usepackage{float}
\usepackage{color}
\usepackage{soul}

\begin{document}

\title{Relevance between Information scrambling and quantum Darwinism}

\author{Feng Tian$^{1}$}
\author{Jian Zou$^{1}$}
\email{zoujian@bit.edu.cn}
\author{Hai Li$^{2}$}
\author{Bin Shao$^{1}$}

\affiliation{$^{1}$School of Physics, Beijing Institute of Technology, Beijing 100081, People's Republic of China}
\affiliation{$^{2}$School of Information and Electronic Engineering, Shandong Technology and Business University, Yantai 264000, People's Republic of China}
\date{Submitted \today}

\begin{abstract}

Quantum system interacting with environment can induce redundant encoding of the information of system into a multipartite environment, which is the essence of quantum Darwinism. At the same time, environment may scramble the initially localized information about the system. We mainly investigate the relevance between information scrambling in environment and the emergence of quantum Darwinism.
First, we generally identify that when the system shows a Darwinistic behavior system information that is initially localized in the environment is not scrambled, while when Darwinism disappears scrambling occurs.
We then verify our result through a collision model where the system, consisting of one or two qubits, interacts with an ensemble of environmental ancillas.
Moreover, dependent on the nature of system-environment interactions, our results also shows that the single qubit and two-qubit systems behave differently for the emergence of QD and the scrambling, but the above relevance between them remains valid.
\end{abstract}

\maketitle

\section{Introduction}
Quantum Darwinism (QD) is a theoretical framework that allows one to understand the emergence of objectivity out of quantum superpositions~\cite{darwin1}, which treats environment as a communication channel.
As the system decoheres due to the interaction with an environment~\cite{decohe1,decohe2,decohe3,decohe4}, selective information of the system---information about the pointer states---is duplicated into different parts of the environment~\cite{darwin2,darwin3,darwin4,darwin5}.
Then many observers can independently access and measure different parts of the environment and independently obtain the same information about the system~\cite{darwin6,darwin7,darwin8}. This redundancy explains the emergence of objective reality.

The emergence of QD has been extensively studied in various models, such as spin system~\cite{theore1}, random unitary model~\cite{theore2}, quantum Brownian~\cite{theore3}, multilevel environment~\cite{theore4}.
One of the main issues of these researches is to understand the fundamental mechanism through which QD emerges.
Recently several works have shown that the emergence of QD is sensitive to the microscopic description of quantum dynamics, such as the nature of interaction and initial conditions~\cite{CMQD1,CMQD2}.
It has been shown that the nature of correlations among the environmental constituents, i.e., whether they are quantum or classical, is important for the emergence of QD~\cite{CMQD2}.
In addition, the relation between QD and non-Markovianity~\cite{nonMarkov1,nonMarkov2,nonMarkov3,nonMarkov4,nonMarkov5,nonMarkov6} have been investigated.
Beyond these, recently some works have elaborated the consistent predictions on objectivity between quantum Darwinism and spectrum broadcasting~\cite{broadcast1,broadcast2,broadcast3,broadcast4,broadcast5,broadcast6}.
Experimental investigations of QD were also reported in photonic cluster ststes~\cite{experiment1}, photonic quantum simulators~\cite{experiment2} and nitrogen vacancy centers~\cite{experiment3}.
Generally QD demands the analysis of quantum mutual information between system and environment, and hence it is necessary to keep track not only of the system but also of the environment.
The master equation is widely used to obtain the dynamics of open system by tracing out environmental degrees of freedom, while it cannot treat the full system-environment dynamics.
The collision model (CM)~\cite{collide1,collide2,collide3,collide4} which decomposes a complicated dynamics in terms of discrete elementary processes, offers an alternative way to the description of open quantum system dynamics. And for CM, the correlations between the open quantum system and its environment can be easily traced.
In the standard framework of CM, the environment is represented by an ensemble of uncorrelated identical environment constituents termed ancillas and the system of interest interacts with each ancilla sequentially. Recently, the CMs have found application in the investigation of QD~\cite{CMQD1,CMQD2,CMQD3}.

In the most general scenario, a quantum system interacts with an environment with a large number of degrees of freedom such that environment forms a many-body system. Very recently, for the characterization of the dynamics in quantum many-body system the question of how quantum information spreads over the constituent degrees of freedom have attracted more and more attention. The delocalization of information in a many body system is referred to as scrambling~\cite{scram1}.
Recently, the study of information scrambling has attracted significant attention in quantum information fields~\cite{scram7,scram17}. Ref.~\cite{scram2} has shown that the scrambling is
 linked with the entanglement propagation in a diffusive system. It has been recognized that the scrambling plays an important role in information propagation~\cite{scram19,scram20}. It is also intimately related to the phenomena of transportation in non-Fermi liquids~\cite{nonFermi}, local thermalization in nonequilibrium many-body system~\cite{scram3}, and black-hole information paradox~\cite{scram4,scram5,scram6}, etc.
A well-known probe of quantum information scrambling is the out-of-time-order correlator (OTOC), whose decay rate is connected to the Lyapunov exponent in the semiclassical limit~\cite{scram7,scram8,scram9}.

Recently, in many-body quantum systems more attention has been paid to tripartite mutual information (TMI), which is an alternative measure of information scrambling. At first, temporal TMI has been investigated by using the channel-state duality in Refs.~\cite{scram10,scram11}. Later, instantaneous TMI also has been used to study information scrambling in Refs.~\cite{scram12,scram13,scram14}.
In terms of TMI, the scrambling of information has been studied in various models, such as quantum lattice~\cite{scram10}, spin-1/2 Ising model~\cite{scram13}, spin chain~\cite{scram14} and collision models~\cite{scram15}. Their typical settings assume that at the initial time, the information of the small system is locally encoded in the many-body system through entanglement. Once the information is encoded, the small system is isolated and does not evolve anymore. The many-body system then evolves unitarily and the locally encoded information will spread over the entire many-body system. Remarkably, different from these configurations, Ref.~\cite{scram16} studied a scenario where a nuclear spin is collectively coupled to a spin bath. Such a difference needs to be emphasized: in this situation, scrambling of information in the bath is not due to the internal dynamics of environment but rather because the spin-bath dynamics take places.

Different from OTOC, TMI is an operator-independent quantity. The method used in this paper is instantaneous TMI, the sign of which indicates whether information is scrambled or not. When TMI is non-negative at some time, the information at this moment is localized, while at some time TMI is negative, the information is delocalized now. If TMI is non-negative at the beginning and becomes negative as time evolves, which means that information turns into delocalized, namely, scrambling occurs.

A full characterization of locally encoded quantum information spread in many-body system could be key in understanding the process of objectivity. This raises the question of what is the interplay between information scrambling and QD.
In this paper, we investigate the emergence of QD and information scrambling from a new perspective, and try to find their interplay. First we give a discussion about their relationship, and arrive at the conclusion that when the system shows a Darwinistic behavior initially localized information is not scrambled in the environment, while when there is no Darwinistic behavior appearing, scrambling occurs.
We then verify the above conclusion by using a CM which consists of a system (containing one qubit or two qubits) and an environment (a quantum-body system) compose of a collection of environment ancillas. In our model, similar to Ref.~\cite{scram16}, the scrambling of initially localized system information in the environment is due to the sequential system-environment interactions, which is different from the typical configurations where the scrambling is because of the internal dynamics of a many-body system~\cite{scram10} or environment~\cite{scram18}.
We consider the pure dephasing and exchange interactions between the system and environment, respectively.
Moreover, dependent on the nature of interactions between the system and environment, our results also shows that the single qubit and two-qubit systems behave differently for the emergence of QD and the scrambling of information, but the relevance between them remains valid.

\begin{figure}[htbp]
\includegraphics[scale=0.45]{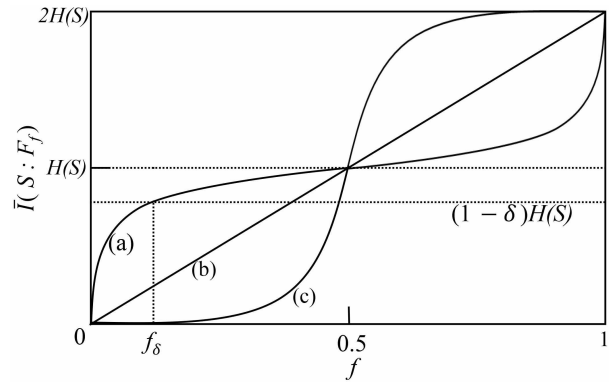}
\centering
\caption{$\bar{I}(S : F_f)$ as a function of $f$. Line (a) corresponds to the signature of QD, where a small fraction $f_\delta$ of $E$ contain almost all (all but $\delta$) of the information about $S$. The linear profile (b) shows the behavior of independent environment. Line (c) shows an ``encoding" environment or an antiredundancy, i.e., $\bar{I}/H_S$ takes on an $S$-shaped profile.}
\label{Fig1}
\end{figure}
\section{Relevance between Quantum Darwinism and information scrambling}\label{Sec2}
In QD the signature of objectivity is described by the quantum mutual information (QMI) between the system $S$ and a fragment of
the environment $F_f \subseteq E$
\begin {equation}\label{eq:1}
I(S:F_f)=H_S+H_{F_f}-H_{SF_f}.
\end {equation}
 Here $H(X)$=$-$Tr$[\rho_{X}\ln\rho_{X}]$ is the von Neumann entropy for the reduced state $\rho_X$ of subsystem $X$, and $f\in[0, 1]$ is the fraction of $E$ contained in $F_f$ (e.g., $F_f$ contains $fN$ individual subsystems if $E$ is compoesd of $N$ individual quantum systems).
The averaged mutual information $\bar{I}(S : F_f)$, is defined by the average of $I(S : F_f)$ over all possible fragments $F_f$ with the same size $f$.
The redundant information encoding throughout the environment is when the averaged mutual information takes value $\bar{I}(S : F_\delta)=(1-\delta)H_S$, i.e., the fragment $F_\delta$ with the size $f_\delta$ is said to contain roughly all the information of the system state.
If this occurs at sufficiently small fragments $F_\delta$, we can say that Darwinistic phenomena emerges. In a plot of $\bar{I}(S : F_f)$ versus fraction size $f$, such redundancy behavior is characterized by a rapid rise of $\bar{I}(S : F_f)$ at relatively small $f$, followed by a long ``classical plateau''. This plateau continue until the fragment of the environment encompasses the whole environment and $\bar{I}$ increases to $2H_S$ [see line (a) in Fig.~\ref{Fig1}].
Line (b) in Fig.~\ref{Fig1} shows that each environment ancilla provides unique and independent information. Line (c) indicates that information about $S$ is encoded in multiple subenvironments and to learn about the system one requires access to at least half of the environmental fraction~\cite{decohe4}.
For pure global state, the curve ($\bar{I}$ vs. $f$) is always antisymmetric with respect to $\bar{I}=H_S$ and $f = 0.5$.

The tripartite mutual information (TMI) has been used as a quantifier of information scrambling. The TMI among three subsystems $A$, $B$, and $C$ is defined as
\begin {equation}\label{eq:2}
I_3(A:B:C)=I(A:B)+I(A:C)-I(A:BC)
\end {equation}
where $I(X:Y)$ is the mutual information between $X$ and $Y$.
From an information-theoretic point of view, TMI quantifies how the total (quantum and classical) information is shared among the subsystems $A$, $B$ and $C$.
TMI is negative when $I(A:B)+I(A:C)<I(A:BC)$, which implies that the sum of the total information that shared between $A$ and $B$, $A$ and $C$ is smaller than that between $A$ and $BC$ together.
In this case, the information about $A$ is nonlocally stored in $B$ and $C$ such that measurements on $B$ and $C$ alone are not able to reconstruct $A$.
Thus, a negative value of TMI is associated with delocalization of the total information, or the total information being scrambled

Now we explore the relation between the emergence of QD and the information scrambling.
Consider a composite system of a quantum system $S$ and an environment $E$ consisting of a collection of $N$ ancillas ($E_1$, $E_2$, ..., $E_N$). At the initial time, information about $S$ is locally encoded in the environment by entangling the system and
the first ancilla $E_1$ in the environment. Specifically, we set the initial system-environment state to be:
\begin {equation}\label{eq:3}
|\Psi_0\rangle=|\phi_{SE_1}\rangle \otimes |\eta_0^1\rangle \otimes |\eta_0^2\rangle \otimes ... |\eta_0^N\rangle.
\end {equation}
Here $|\phi_{SE_1}\rangle$ is an entangled state between the system and ancilla $E_1$, and $|\eta_0^j\rangle$ $(j=1, 2, ..., N)$ is the initial state of $E_j$, respectively.
Dependent on different system-environment dynamics, the joint system-environment state after the evolution is also different.
We first assume that the composite system, after time $t$, evolves into the state,
\begin {equation}\label{eq:4}
|\Psi_t\rangle= \sum\limits_{k} \alpha_k|\phi_k\rangle \otimes |\eta_k^1\rangle \otimes |\eta_k^2\rangle \otimes ... |\eta_k^N\rangle
\end {equation}
where $|\phi_k\rangle$ is the pointer states of the system and $\{|\eta_k^j\rangle\}$ being the eigenbasis of the subenviornment $E_j$. In fact, the state of Eq.~(\ref{eq:4}) is a extreme case where QD emerges perfectly in the sense that even a single environmental ancilla is sufficient to access all the information about $S$.
To study how it relates to the information scrambling measured by the TMI, we divide the environment $E$ into three nonoverlapping subsystems $E_1\in B$, $\{E_2, ..., E_{m+2}\}\in C$, and $\{E_{m+3}, ..., E_N\}\in D$.
We now consider the mutual information between $S$ and $B$($C$). In terms of Eq.~(\ref{eq:4}), tracing out all subsystems that are not associated with $S$, we obtain the reduced density matrix of the system $S$
\begin {equation}\label{eq:5}
\rho_S= \sum\limits_{k} |\alpha_k|^2  |\phi_k\rangle\langle\phi_k|
\end {equation}

Similarly, we can obtain
\begin {equation}\label{eq:6}
\rho_{B(C)}= \sum\limits_{k} |\alpha_k|^2 |\varphi_k^{B(C)}\rangle\langle\varphi_k^{B(C)}|,
\end {equation}
\begin {equation}\label{eq:7}
\rho_{SB(SC)}= \sum\limits_{k} |\alpha_k|^2  |\phi_k\rangle\langle\phi_k|   \otimes  |\varphi_k^{B(C)}\rangle\langle\varphi_k^{B(C)}|,
\end {equation}
where $\varphi_k^{B} :=\otimes_{j\in B} |\eta_k^j\rangle$ and $\varphi_k^{C} :=\otimes_{j\in C} |\eta_k^j\rangle$.
Using the definition of Eq.~(\ref{eq:1}), we have
\begin {equation}\label{eq:8}
I(S:B)=I(S:C)=H(S)=-\sum\limits_{k}|\alpha_k|^2\log_2p_k,
\end {equation}
which is independent of $m$. This means that from the joint system and environment state of Eq.~(\ref{eq:4}), one can acquire information about the system through the measurement on any size of the fraction of the environment. In other words, the state of Eq.~(\ref{eq:4}) perfectly exhibits the emergence of QD. According to Eq.~(\ref{eq:2}) the TMI among the $S$, $B$ and $C$ can be obtained as
\begin {equation}\label{eq:9}
I_3(S:B:C) = -\sum\limits_{k}|\alpha_k|^2\log_2p_k>0.
\end {equation}
This positive TMI indicates that when QD emerges, initially localized information about the system is not scrambled in the environment. This happens because, in this case more information about the system $S$ is shared among individual subenvironment, and thus measurement on a smaller fraction of the environment can have access to almost the same amount of information about the system, leading to the emergence of QD.

\begin{figure}[htbp]
\flushleft
\includegraphics[scale=0.5]{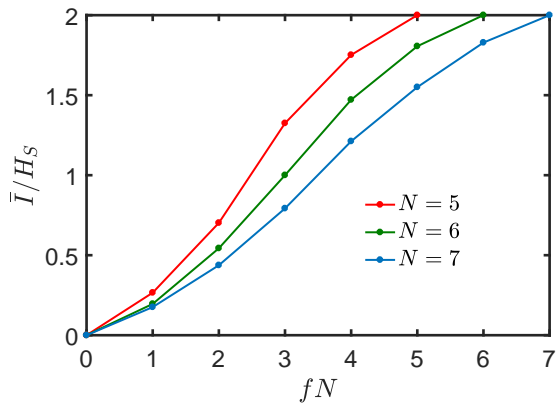}
\centering
\caption{Normalized mutual information $\bar{I}/H_S$ as a function of $fN$ for different $N$. The plotted curves are the results of averaging over 1000 possible environment fragments with the same $f$.}
\label{Fig2}
\end{figure}

Now we consider another case where no QD appearing. For the same initial state Eq.~(\ref{eq:3}), we assume that, after time $t$, it evolves into the state,
\begin {equation}\label{eq:10}
|\Phi_t\rangle= \sum\limits_{k} \alpha_k|\phi_k\rangle \otimes |\xi_k\rangle,
\end {equation}
where $|\xi_k\rangle$ are entangled environment states. In this case, due to the presence of entanglement between the subenvironments, information about the system is shared among the joint environment ancillas rather than with the individual ancilla. Hence, any measurement on a smaller fraction of the environment may not obtain enough information about the system. To illustrate this, assume the main system $S$ is a qubit and the environment is composed of $N$ qubits, for which we assume that the evolved state $|\Phi_t\rangle$ is
\begin {equation}\label{eq:11}
|\Phi_t\rangle= |0\rangle \otimes |D_N^{(2)}\rangle + |1\rangle \otimes |D_N^{(1)}\rangle,
\end {equation}
where environmental state $|D_N^{(d)}\rangle$ is $N$-qubit Dicke state with $d$ excitations, defined as
\begin {equation}\label{eq:12}
|D_N^{(d)}\rangle=\binom{N}{d}^{-1/2} \sum\limits_{i}P_i \{|1\rangle^{\otimes d} |0\rangle^{\otimes {(N-d)}} \},
\end {equation}
where $\sum\limits_{i}P_i\{\cdot\}$ denotes the sum over all possible permutations. From Eq.~(\ref{eq:11}), in Fig.~\ref{Fig2} we plot normalized mutual information $\bar{I}/H_S$ as a function of environment fractions $fN$ for different environment sizes $N=5$, $N=6$ and $N=7$. It can be seen that $\bar{I}/H_S$ takes on an $S$-shaped profile indicating an ``encoding" environment.
To investigate its links with the scrambling, we then calculate the TMI between the $S$, $B$ and $C$ in this case, and obtain $I_3(S:B:C)<0$
independent of the choice of $m$. This negative TMI implies that when there is no any signatures of QD appearing initially localized system information is scrambled in the environment. This can be understood as following: in this case initially localized information is predominantly nonlocally shared among the joint system $BC$, which cannot be detected by local measurement just on $B$ or $C$. In other words, measurement just on a smaller fraction of the environment cannot acquire the information of system state.
In the next section, we will explore their relationship quantitatively from a more specific model.

\section{model}\label{Sec3}
\begin{figure}[htbp]
\includegraphics[scale=0.32]{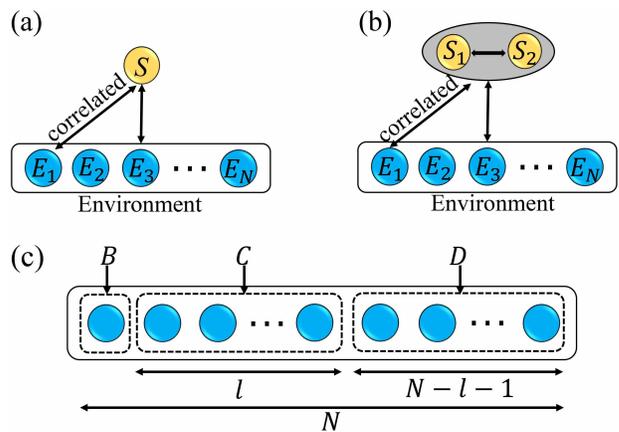}
\centering
\caption{Schematics of the collisional model. (a) A single qubit system $S$ and (b) two interacting qubit system $S$ couple to a environment $E$ composed of a collection of ancillas ($E_1, E_2, ..., E_N$), respectively. For both cases, the ancilla $E_1$ is initially correlated with $S$, and the left ancillas $E_k (k\geq2)$ are initialized in the same state  $|\eta_k\rangle$. (c) The whole environment $E$ is divided into three nonoverlapping subsystems $B$, $C$, $D$, whose sizes (the numbers of the ancillas) are, respectively, given by 1, $l$, and $N-l-1$.}
\label{Fig3}
\end{figure}
We consider a collision model, which consists of a system $S$ and an environment $E$. The environment consists of a collection of $N$ non-interacting environment ancillas $(E_1,E_2,...,E_N)$, which is considered as a many-body system. We study two cases where the system $S$ contains a single qubit (Fig.~\ref{Fig3}(a)) and two qubits (Fig.~\ref{Fig3}(b)), respectively. For the single qubit system, the Hamiltonian is given by
\begin {equation}\label{eq:13}
{H}_{S}^1=\frac{1}{2} {\sigma}_z
\end {equation}
with $z$-Pauli operator ${\sigma}_z$ (we set $\hbar = 1$).
For the two-qubit system, the Hamiltonian is
\begin {equation}\label{eq:14}
{H}_{S}^2 = {H}_{S_1}+{H}_{S_2}+{H}_{S_1,S_2},
\end {equation}
where ${H}_{S_i }=\frac{1}{2} {\sigma}_z (i=1,2)$ is the free Hamiltonian of $S_i$ and
\begin {equation}\label{eq:15}
{H}_{S_1,S_2}=\epsilon({\sigma}_x\otimes{\sigma}_x + {\sigma}_y\otimes{\sigma}_y)
\end {equation}
is the interaction Hamiltonian between $S_1$ and $S_2$ with coupling strength $\epsilon$.
We assume throughout this paper that each ancilla of the environment is qubit, and its Hamiltonian is given by
${H}_{S,E_k}=\frac{1}{2} {\sigma}_z$.

Initially, information of the system is locally encoded in the environment by entangling the system and the first ancilla $E_1$ in the environment. Specifically, we set the initial state of the system and environmental ancillas to be:
\begin {equation}\label{eq:16}
|\psi_0\rangle=|\phi_{SE_1}\rangle  \mathop{\otimes}\limits_{k=2}^{N} |\eta_k\rangle,
\end {equation}
where $|\phi_{SE_1}\rangle$ is the entangled state between the system and ancilla $E_1$, and $|\eta_k\rangle$ are the initial state of the $k$th $(k\geq2)$ ancilla $E_k$.
A schematic sketch of the collision model is given in Fig.~\ref{Fig3}.

We consider the general Heisenberg interaction between system and ancilla:
\begin {equation}\label{eq:15}
{H}_{S,E_k}=\mathop{\sum}\limits_{j=x,y,z} J_j(\sigma_S^j \otimes \sigma_{E_k}^j),
\end {equation}
with coupling strength $J_j\ (j=x,y,z)$.
In any $n$th step, the $S-E_k$ interaction is realized by the application of the unitary operation
\begin {equation}\label{eq:16}
U_{S,E_k}=\exp{[-i({H}_{0}+{H}_{S,E_k})t]},
\end {equation}
where ${H}_{0}=H_S^{1(2)}+H_{E_k}$ for the single (two-qubit) system, and $t$ stands for the interaction time, i.e., the duration of each collision. Thus the joint system-environment state $\rho_0=|\psi_0\rangle\langle\psi_0|$, after $N$ steps, evolves into the state
\begin {equation}\label{eq:17}
\rho_N=U_{\{N\}}\rho_0 U_{\{N\}}^\dag
\end {equation}
with $U_{{\{N\}}}=U_{S,E_{N}}U_{S,E_{N-1}}...U_{S,E_2}U_{S,E_1}$.

To investigate the information scrambling, we divide the whole environment $E$ into three nonoverlapping subsystems $B$, $C$, $D$, whose sizes (the number of the ancillas) are, respectively, given by 1, $l$, and $N-l-1$, as shown in Fig.~\ref{Fig3}(c). Then we compute the averaged TMI $\bar{I}_3(S:B:C)$ defined as the averaging over possible partitions of $E$ for a given $l$.
\section{results}\label{Sec5}
In this paper we will examine two types of system-environment interactions, the pure dephasing interaction and exchange interaction, for both the single qubit and two-qubit systems, repectively.
\subsection{Dephasing channel}
\begin{figure}[htbp]
\flushleft
\includegraphics[scale=0.6]{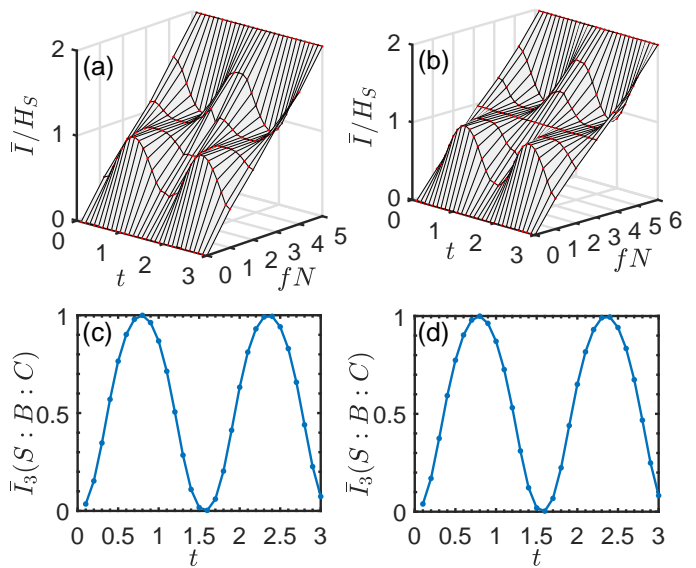}
\centering
\caption{Normalized mutual information $\bar{I}/H_S$ and averaged TMI $\bar{I}_3$ for the pure dephasing interaction in the case of single qubit system. Upper panel: $\bar{I}/H_S$ versus $fN$ and $t$. Lower panel: $\bar{I}_3(S:B:C)$ as a function of $t$. (a), (c) $N=5$. (b), (d) $N=6$. We set $\epsilon=1$, $J=1$ and $l=2$. The plotted curves are the results of averaging over 1000 possible environment fragments with the same $f$ for $\bar{I}/H_S$ and 1000 possible environment partitions for $\bar{I}_3$, respectively.
}
\label{Fig4}
\end{figure}
\begin{figure}[htbp]
\flushleft
\includegraphics[scale=0.5]{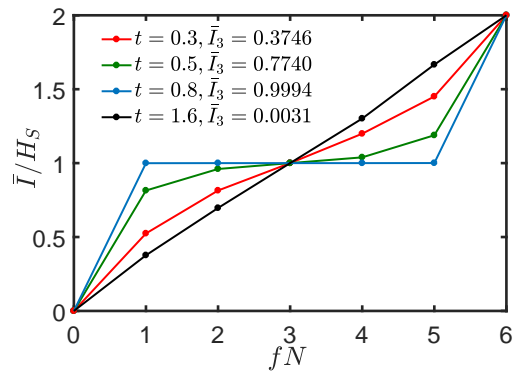}
\centering
\caption{Normalized mutual information $\bar{I}/H_S$ as a function of $fN$ at different $t$. All parameters are the same as those in Fig.~\ref{Fig4}(b).}
\label{Fig5}
\end{figure}
First we examine the pure dephasing channel, i.e., $J_x=J_y=0$ and $J_z=J$. First we consider the single qubit system.
The initial entangled state $|\phi_{SE_1}\rangle$ between the system and ancilla $E_1$ is prepared in $(|-+\rangle+|+-\rangle)/\sqrt{2}$ with $|\pm\rangle=(|0\rangle\pm|1\rangle)/\sqrt{2}$. And all the ancillas $E_k$ ($k\geq2$) are initially prepared in identical state $|\eta_k\rangle=(|0\rangle+|1\rangle)/\sqrt{2}$.
For the single qubit system, we show in Figs.~\ref{Fig4}(a) and 4(b) the normalized mutual information $\bar{I}/H_S$ versus time $t$ and environment fractions $fN$. To understand the connection between QD and information scrambling, we also plot the corresponding averaged $\bar{I}_3(S:B:C)$ as a function of $t$ in Figs.~\ref{Fig4}(c) and 4(d), respectively. It can be seen from Figs.~\ref{Fig4}(a) and 4(b) that at small $t$, the mutual information increases almost linearly with the size of the environment fragment. As $t$ increases, first we can see that a small fraction of the environment already has averaged mutual information close to $H(S)$, being the signature of QD. That is, the redundancy plateau gets more and more pronounced with the increase of $t$, and then a sharp characteristic plateau appears.
Subsequently, this clear signatures of objectivity is gradually lost and the redundancy plateau emerges periodically as $t$ increases further. From Figs.~\ref{Fig4}(c) and 4(d), it can be seen that $\bar{I}_3(S:B:C)$ is positive, and it experiences successive increasing and decreasing behaviors with the increase of $t$. We find, quite interestingly, the change of $\bar{I}_3$ can be correlated to the emergence of objectivity as witnessed by QD. Specifically, when the positive $\bar{I}_3$ increases, the plateau gets more pronounced. And the sharp plateau occurs when the positive $\bar{I}_3$ is maximized. The redundancy plateau is suppressed when the positive $\bar{I}_3$ decreases. To clearly illustrate this, in Fig.~\ref{Fig5} we plot $\bar{I}/H_S$ as a function of $fN$ at some instants of time corresponding to different value of $\bar{I}_3(S:B:C)$.
From Fig.~\ref{Fig5} we find that, for a small $\bar{I}_3$ ($\bar{I}_3=0.0031$), the mutual information varies almost linearly with the size of the environment fragments (black line). It is clear that as $\bar{I}_3$ grows, the signature of QD begins to appear and become apparent (see the red ($\bar{I}_3=0.3746$) and green ($\bar{I}_3=0.7740$ ) lines. For a large $\bar{I}_3$ ($\bar{I}_3=0.9994$), the sharp characteristic plateau appears (blue line) such that even a single ancilla give essentially all classical information about $S$ and additional ancillas just confirm what is already known. Physically, this can be understood as following: a larger positive value of $\bar{I}_3$ indicates that more information about the system $S$ is shared among individual environment ancillas. Thus, measurement on a smaller fraction of the environment can have access to almost the same amount of information about the system, leading to the emergence of QD. In contrast, a smaller positive value of $\bar{I}_3$ means that measurement on an individual ancilla knows less information about the system such that a larger fraction of the environment is required to know the information about the system state.
\begin{figure}[htbp]
\flushleft
\includegraphics[scale=0.6]{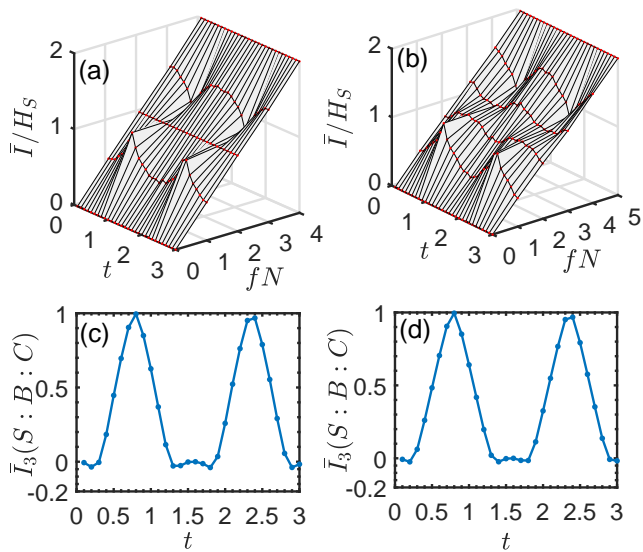}
\centering
\caption{Normalized mutual information $\bar{I}/H_S$ and averaged TMI $\bar{I}_3$ for the pure dephasing interaction in the case of two-qubit system. Upper panel: $\bar{I}/H_S$ versus $fN$ and $t$. Lower panel: $\bar{I}_3(S:B:C)$ as a function of $t$. (a), (c) $N=4$. (b), (d) $N=5$. We set $\epsilon=1$, $J=1$ and $l=2$. The plotted curves are the results of averaging over 1000 possible environment fragments with the same $f$ for $\bar{I}/H_S$ and 1000 possible environment partitions for $\bar{I}_3$, respectively. }
\label{Fig6}
\end{figure}
\begin{figure}[t]
\flushleft
\includegraphics[scale=0.6]{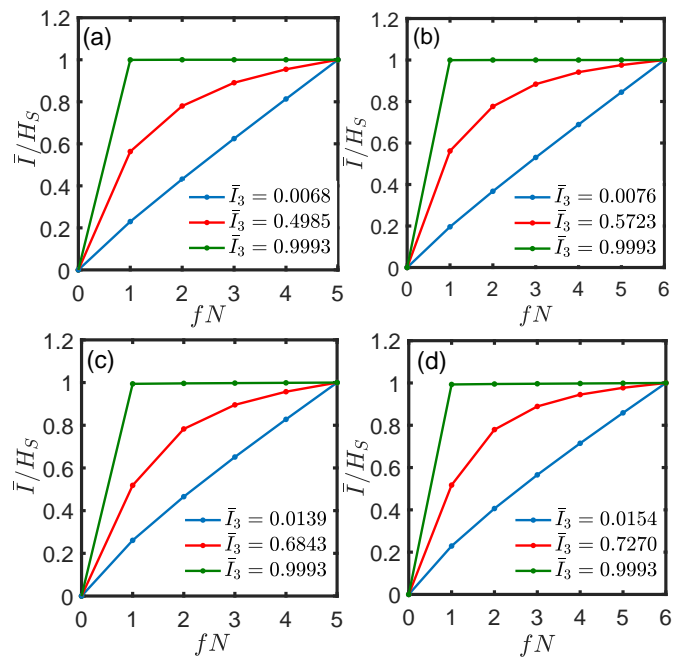}
\centering
\caption{Normalized mutual information $\bar{I}/H_S$ as a function of $fN$ for the pure dephasing interaction. Upper panel: single qubit system with the same parameters as in Fig.~\ref{Fig4}. Lower panel: two-qubit system with the same parameters as in Fig.~\ref{Fig6}. The blue, red, and green lines correspond to $t$=0.05, 0.8 and 1.2, respectively. (a), (c) $N=10$. (b), (d) $N=12$.}
\label{Fig7}
\end{figure}

Now we consider the two-qubit system.
The initial entangled state $|\phi_{SE_1}\rangle$ between the system and ancilla $E_1$ is prepared in $(|-+-\rangle+|+-+\rangle)/\sqrt{2}$, and all the ancillas $E_k$ ($k\geq2$) are initially prepared in identical state $|\eta_k\rangle=(|0\rangle+|1\rangle)/\sqrt{2}$.
Figs.~\ref{Fig6}(a) and 6(b) show the normalized mutual information $\bar{I}/H_S$ varying with $t$ and $fN$. In this case, in spite of a quantitative difference, its behaviors are similar to those of single qubit system. We also see that a redundancy plateau emerges periodically with the increase of $t$. We also examining the behavior of the $\bar{I}_3(t)$ in Figs.~\ref{Fig6}(c) and 6(d). And we again arrive at a conclusion similar to that of single qubit system: the larger positive value of $\bar{I}_3$, the more pronounced the Darwinistic plateau gets and vice versa.

Until now, we have only considered the environment $E$ consisting of up to $N=6$ ancillas. In Fig.~\ref{Fig7} we show that above effects persist for larger environment when $N=10$ and $N=12$. Because the global $S+E$ state is pure, the plots ($\bar{I}/H_S$ versus~$f$) is antisymmetric about $f=1/2$. Based on this, for convenience of calculations, we plot $\bar{I}/H_S$ as a function of $fN$ varying from 0 to $N/2$. It can be seen from Fig.~\ref{Fig7} that the larger positive TMI, the more pronounced the Darwinistic plateau gets and
vice versa, showing that above results are valid for larger environments.
\subsection{Exchange interaction}
\begin{figure}[h]
\flushleft
\includegraphics[scale=0.6]{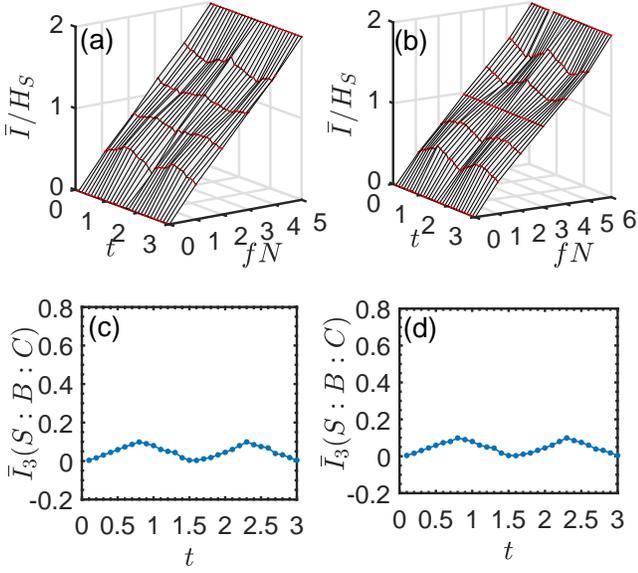}
\centering
\caption{Normalized mutual information $\bar{I}/H_S$ and averaged TMI $\bar{I}_3$ for the exchange interaction in the case of single qubit system. Upper panel: $\bar{I}/H_S$ versus $fN$ and $t$. Lower panel: $\bar{I}_3(S:B:C)$ as a function of $t$. (a), (c) $N=5$. (b), (d) $N=6$. We set $\epsilon=1$, $J=1$ and $l=2$. The plotted curves are the results of averaging over 1000 possible environment fragments with the same $f$ for $\bar{I}/H_S$ and 1000 possible environment partitions for $\bar{I}_3$, respectively. }
\label{Fig8}
\end{figure}
\begin{figure}[htbp]
\flushleft
\includegraphics[scale=0.6]{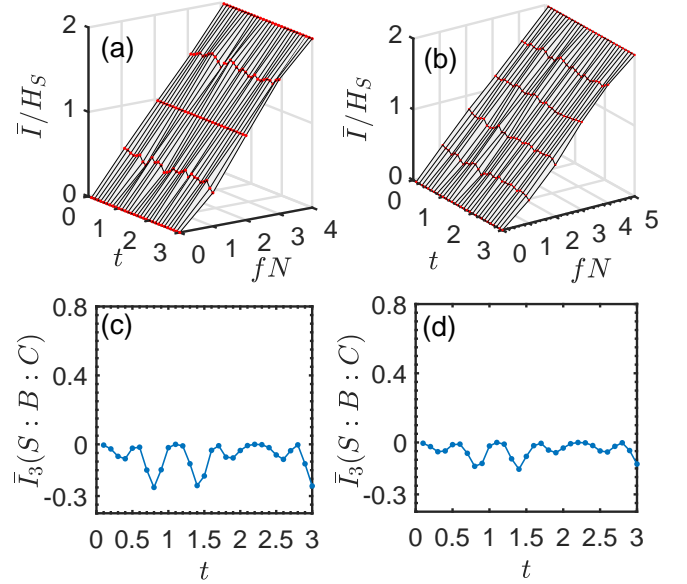}
\centering
\caption{Normalized mutual information $\bar{I}/H_S$ and averaged TMI $\bar{I}_3$ for the exchange interaction in the case of two-qubit system. Upper panel: $\bar{I}/H_S$ versus $fN$ and $t$. Lower panel: $\bar{I}_3(S:B:C)$ as a function of $t$. (a), (c) $N=5$. (b), (d) $N=6$. We set $\epsilon=1$, $J=1$ and $l=2$. The plotted curves are the results of averaging over 1000 possible environment fragments with the same $f$ for $\bar{I}/H_S$ and 1000 possible environment partitions for $\bar{I}_3$, respectively.}
\label{Fig9}
\end{figure}
Now we turn our attention to the exchange interaction, i.e., $J_x=J_y=J$ and $J_z=0$. First we consider the single qubit system.
The initial entangled state $|\phi_{SE_1}\rangle$ between the system and ancilla $E_1$ is initially prepared in the same state as that in Sec.~\ref{Sec5}A, while all the ancillas $E_k$ are initially prepared in identical state $|\eta_k\rangle=|0\rangle$.
For the single qubit system, in Figs.~\ref{Fig8}(a) and 8(b) we show the normalized mutual information $\bar{I}/H_S$ varying with $fN$ and $t$. In this case, there is no sharp plateau appearing when compared to the case of the pure dephasing interaction. Nonetheless, the key features of quantum Darwinism still persists, i.e., the functional behavior of $\bar{I}/H_S$ is still qualitatively consistent with QD. We then examining the corresponding behavior of $\bar{I}_3$ in Figs.~\ref{Fig8}(c) and 8(d). It can be seen from Figs.~\ref{Fig8}(c) and 8(d) that $\bar{I}_3$ is much smaller compared with that in Figs.~\ref{Fig4}(c) and 4(d) for all time.
This indicates that less quantum information about the system $S$ is shared among individual environment ancillas. Thus, measurement on a smaller fraction of the environment can only have access to less information about the system. This is why the signature of QD is less manifest in this case.

\begin{figure}[htbp]
\flushleft
\includegraphics[scale=0.5]{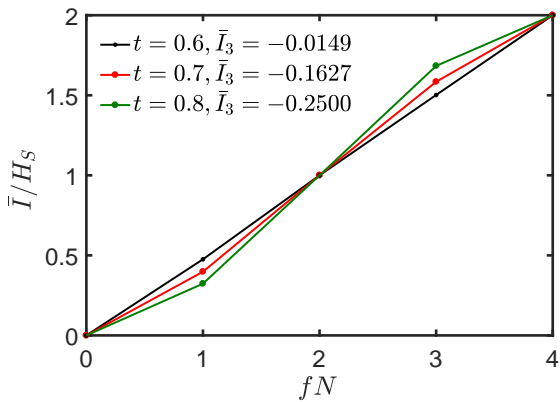}
\centering
\caption{Normalized mutual information $\bar{I}/H_S$ as a function of $fN$ at different $t$. All parameters are the same as those in Fig.~\ref{Fig9}(a).}
\label{Fig10}
\end{figure}

Now we consider the two-qubit system.
The initial entangled state $|\phi_{SE_1}\rangle$ between the system and ancilla $E_1$ is initially prepared in the same state as that in Sec.~\ref{Sec5}A, while all the ancillas $E_k$ are initially prepared in identical state $|\eta_k\rangle=|0\rangle$. Figs.~\ref{Fig9}(a) and 9(b) show the normalized mutual information $\bar{I}/H_S$ as a function of $t$ and $fN$. Different from the single qubit system, we find that the signature of QD is completely lost and there is no redundant encoding of system information. And at some time-window, we find that $\bar{I}/H_S$ takes on an $S$-shaped profile indicating an ``encoding" environment or an antiredundancy, i.e., information about $S$ is encoded in the multiple ancillas and to learn about the system one requires access to at least half of the environmental ancillas.
To understand the emergence of this antiredundancy behavior, we examine the corresponding $\bar{I}_3$ in Figs.~\ref{Fig9}(c) and 9(d).
Interestingly, it can be seen that when $\bar{I}/H_S$ takes on an $S$-shaped profile, $\bar{I}_3$ becomes negative indicating the scrambling or delocalization of quantum information. Specifically, a negative $\bar{I}_3$ with a larger absolute value corresponds to a stronger antiredundancy and vice versa. To clearly illustrate this, in Fig.~\ref{Fig10} we plot the normalized mutual information $\bar{I}/H_S$ as a function of $fN$ at some instants corresponding different value of $\bar{I}_3$. It can be seen in Fig.~\ref{Fig10} that $\bar{I}_3$ with a larger absolute value corresponds to a stronger antiredundancy (i.e., the more pronounced $S$-shaped profile of the curve). This happens because, a negative $\bar{I}_3$ with a larger absolute value means that more information about the system $S$ is nonlocally stored in the joint system $BC$ and cannot be detected by local measurement just on $B$ or $C$. Thus, we conclude that the delocalization of quantum information play detrimental role in the emergence of QD feature.

\begin{figure}[htbp]
\flushleft
\includegraphics[scale=0.55]{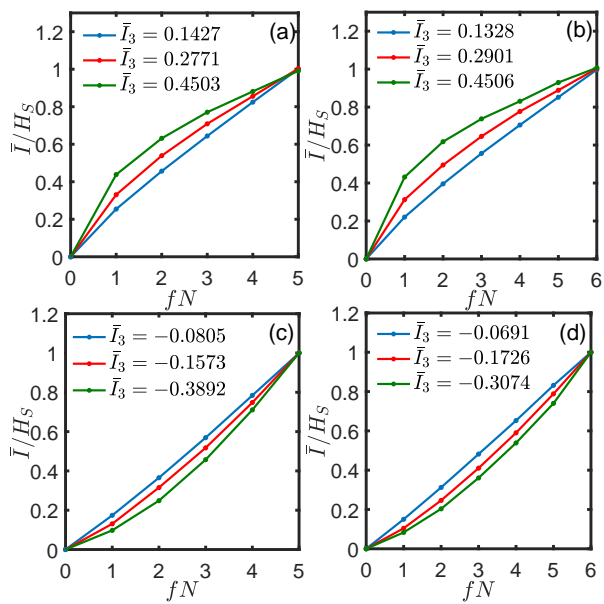}
\centering
\caption{Normalized mutual information $\bar{I}/H_S$ as a function of $fN$ for the exchange interaction. Upper panel: single qubit system with the same parameters as in Fig.~\ref{Fig8}. Lower panel: two-qubit system with the same parameters as in Fig.~\ref{Fig9}. The blue, red, and green lines correspond to $t$=0.3, 0.5 and 0.8, respectively. (a), (c) $N=10$. (b), (d) $N=12$.}
\label{Fig11}
\end{figure}

We also show that above results persist for larger environments with $N=10$ and $N=12$ in Fig.~\ref{Fig11} where $fN$ varies from 0 to $N/2$. For the single qubit system in Figs.~\ref{Fig11}(a) and 11(b), when $\bar{I}_3>0$, as the $\bar{I}_3$ increases redundant features of QD gets more pronounced. For the two-qubit system in Figs.~\ref{Fig11}(c) and 10(d), when information is scrambled ($\bar{I}_3<0$) the features of QD is completely lost. And $\bar{I}_3$ with a larger absolute value corresponds to a stronger antiredundancy.

Besides, dependent on the nature of interactions between the system and environment, our results also shows that the single qubit and two-qubit systems behave differently for the emergence of QD and scrambling. For the dephasing interaction, both of them allow the redundant encoding to emerge. Yet, this is not for the exchange interaction where the redundant encoding is completely lost for the two-qubit system.

\section{Conclusions}
We mainly address the relation between the emergence of QD and information scrambling in the environment. At first, we generally illustrate a correlation between information scrambling and the emergence of Darwinistic behavior.
Our main effort shows that when the system shows a Darwinistic behavior initially localized system information is not scrambled in the environment, while when Darwinism disappears scrambling occurs. This is because: when information is scrambled in the environment, initially localized system information is predominantly nonlocally shared among the environment ancillas. In this case, local measurement just on a smaller fraction of the environment cannot acquire the information of system, leading the lost of QD feature. In contrast, when there is no scrambling appearing, more information is shared among individual environment subsystems. Thus, measurement on a smaller fraction of the environment can have access to almost the same amount of information about the system, leading to the emergence of QD.

Then we  verify our result through a collision model which consists of a system interacting with an ensemble of environment ancillas.
We study two cases where the system contains a single qubit and two qubits, respectively, and examine two types of system-ancilla interactions, the pure dephasing interaction and exchange interaction. For the pure dephasing interaction, we find that QD emerges for both the single qubit and two-qubit systems. And the corresponding TMI have a positive value. Specifically, the larger positive value of TMI, the more pronounced the Darwinistic plateau gets and vice versa. For the exchange interaction, we find that the single qubit and two-qubit systems behave differently for the emergence of QD and scrambling. For the single qubit system, although no sharp plateau appearing, the key features of quantum Darwinism still persists. And we find that this behavior coincides with small but positive TMI. However, for the two-qubit system, there is no redundant encoding of the system information and we do not see any signatures of QD. And we find that the corresponding TMI is negative, indicating that information initially localized in the system is scrambled in the environment. Specifically, the larger negative value of TMI, the stronger antiredundancy.

\section{acknowledgement}
This work is supported by the National Natural Science Foundation of China (Grant Nos.~11775019 and 11875086).

\end{document}